\documentclass[aps,preprint,showpacs,byrevtex]{revtex4}

\newcommand{\be}{\begin{equation}}
\newcommand{\ee}{\end{equation}}
\newcommand{\bea}{\begin{eqnarray}}
\newcommand{\eea}{\end{eqnarray}}
\renewcommand{\d}{\text{d}}

\usepackage{graphics,amsfonts}
\begin{document}

\title{Zeta Function Zeros, Powers of Primes, and Quantum Chaos} 

\author{Jamal Sakhr}
\author{Rajat K. Bhaduri}
\author{Brandon P. van Zyl}
\affiliation{Department of Physics and Astronomy, McMaster University,
Hamilton, Ontario, Canada L8S~4M1}

\date{\today}

\begin{abstract}
We present a numerical study of Riemann's formula for the oscillating 
part of the density of the primes and their powers. The formula is 
comprised of an infinite series of oscillatory terms, 
one for each zero of the 
zeta function on the critical line and was derived by 
Riemann in his paper on primes assuming the Riemann hypothesis.
We show that high resolution spectral lines can be generated   
by the truncated series at all powers of primes and demonstrate explicitly 
that the relative line intensities are correct. We then derive a 
Gaussian sum rule for Riemann's formula. 
This is used to analyze the numerical convergence of the truncated series.  
The connections to quantum chaos and semiclassical physics are discussed. 
\end{abstract}


\maketitle

\section{Introduction}

The Riemann zeta function is at the heart of number theory, and has also
played a pivotal role in the study of quantum chaos \cite{stockmann}.  
There is a deep connection between the complex zeros of the zeta function 
and Random Matrix Theory \cite{snaith}. 
The zeros possess the same statistical properties as 
the energy eigenvalues of a dynamical Hamiltonian that is nonintegrable and  
whose dynamics are not time-reversal invariant. Unfortunately, 
this Hamiltonian is not known in terms of its dynamical variables. 
The main source of insight into this unknown quantum chaotic 
system comes from Gutzwiller's pioneering work \cite{gutz2}, that     
expresses the oscillatory part of the quantum density of states as a sum 
over classical periodic orbits. (Such sums are now referred to as 
trace formulas.) It is well-known that 
the oscillatory part of the density of the Riemann zeros is given 
by a Gutzwiller-like sum, with one periodic term for every integer power 
of a prime number \cite{berry}. 
(A smoothed density of the Riemann zeros has also been studied in 
Ref.~\cite{bhaduri}.) From this perspective, one can infer that a 
spectrum consisting of the Riemann zeros is generated by a 
Hamiltonian (albeit unknown) 
whose classical orbits have actions that are logarithms of primes and integer 
powers of primes.

Conversely, one could ask whether it is also possible 
to calculate the prime number sequence from a
sum of oscillatory terms, with one term for every zero of the zeta function. 
Although less widely-known, such a series was actually given by Riemann 
himself \cite{riemann}. 
Riemann derived an exact formula for the density of the primes (and their 
integer powers) that can be 
expressed as the sum of a smooth function and an infinite 
series of oscillatory terms 
involving the complex zeros of the zeta function. The smooth part has been 
thoroughly studied in the context of the prime number theorem whereas the 
oscillatory part has been largely ignored. Interestingly, it is the latter 
that contains the essential information about the location of the primes, 
as shown below.  
There is a vast literature on the distribution of the prime 
numbers. It is recognized that their distribution exhibits global regularity 
and local irregularity \cite{tenebaum}. The nearest-neighbour spacings 
(NNS) of the primes is known to be Poisson-like \cite{porter}, 
corresponding to an almost uncorrelated random distribution. 
This is very different in character 
from the Gaussian Unitary Ensemble (GUE) distribution of the Riemann zeros. 
Nevertheless, 
it is possible to generate the almost-uncorrelated sequence of 
all integer powers 
of primes from the interference of the highly-correlated Riemann zeros.  

As mentioned above, from the perspective of 
semiclassical periodic orbit theory, 
the density of the Riemann zeros has the structure of  
a dynamical trace formula with periodic orbits. It is natural to ask 
whether Riemann's formula is a trace formula for the primes. 
Despite having the oscillatory terms, as discussed below, Riemann's formula 
is not a trace formula of dynamical origin. 
But, this does not preclude the existence of a trace formula for 
the primes. If one does exist, then this would support the notion that 
there exists a Hamiltonian system whose quantum spectrum is the primes. 
In any case, the exclusion of Riemann's formula as a trace formula 
suggests that there would be no correspondence between the classical 
dynamics and the Riemann zeros for this system. 

The purpose of this paper is to study the density of the 
primes from the perspective of numerical semiclassics.  
To our knowledge, Riemann's series has not been studied numerically.
We first verify that Riemann's formula 
does produce spectral lines at the positions of the primes and 
their powers, even when the series is truncated. 
This is not completely unexpected 
since Riemann's series converges conditionally to the exact 
density which is a set of $\delta$-function spikes. However, the 
$\delta$ functions arise from the \emph{entire} series. The truncated series 
is an approximation to the exact density.
It does not yield spikes, but rather lines of various widths, heights and 
(unknown) shapes and it is not at all obvious 
that the relative line intensities of the truncated series  
are correct. We examine this problem both numerically and analytically.  
We then provide a simple rule for estimating the value of the largest zero 
required to sufficiently resolve individual lines 
of a specific shape in some 
interval of interest and describe how to control the error 
incurred from a truncation of the series.  
 
\section{Riemann's Formula}

We start from the Euler product formula
\be
\zeta(\beta)=\prod_p(1-p^{-\beta})^{-1}, \quad \quad \text{Re}~ \beta > 1,
\ee
where the product is over all primes $p$. It follows that
$\sum_p \sum_{n=1}^{\infty}~ {1 \over n} \exp(-n \beta \ln p) = 
\ln \zeta(\beta)$. Dividing both sides by $\beta$ and then taking the 
inverse Laplace transform of both sides with respect to $E$, we
immediately obtain
\be \label{best1}
N(E)=\sum_p \sum_n~{1 \over n} \Theta \left(E-\ln p^n\right) = 
{1 \over {2 \pi i}}
\int_{a-i \infty}^{a+i \infty} {\ln \zeta(\beta) \over \beta}
e^{\beta E} \d \beta,
\ee
where $a>1$. Riemann evaluated the RHS of Eq.~(\ref{best1}) to
obtain $N(E)$. Upon differentiation with respect to $E$ and the 
subsequent substitution $x=e^E$, we obtain the density $\rho(x)$ of $p^n/n$
along the real axis $x$ as
\be \label{best2}
\rho(x)={1 \over {\ln x}}-{1 \over {x\left(x^2-1\right)\ln x}}
-2 \sum_{\alpha > 0}{\cos(\alpha \ln x) \over {x^{1/2}\ln x}},
\ee
where $x>1$. 
This formula assumes the Riemann hypothesis, which states that the 
infinite number of complex zeros of the zeta function all lie on the
critical line $\beta = (1/2 \pm i \alpha)$, where $\alpha$ is real and 
positive. Note that explicit use of the symmetry of the complex zeros  
has been
made to reduce the summation to cosine functions. 
A generalized version of the Riemann formula, where the zeros 
may lie anywhere in the critical strip is given in Ref.~\cite{keating}.
We shall denote the sum over the oscillatory
terms on the RHS of Eq.~(\ref{best2}) as $\tilde{\rho}(x)$. Since 
Eq.~(\ref{best2}) is exact, it is clear that the $\delta$-function 
spikes of $\rho(x)$  must be generated 
from the interference of the terms in $\tilde{\rho}(x)$.  
From our experience in periodic orbit quantization \cite{brack}, 
we know that a coarse-grained version of the exact density 
of states can be reproduced even from a truncated periodic orbit sum. 
Therefore, in the following section, we focus on the numerical analysis  
of the truncated series.

Before presenting the results, however, we briefly review the pioneering 
numerical work of Riesel and G\"{o}hl~\cite{riesel}.  
The LHS of Eq.~(\ref{best1}) is a set of step functions,
with unit steps at every prime $p$, one-half steps at $p^2$, one-third steps 
at $p^3$ and so on, and may be obtained by taking the contour integral of 
$\ln \zeta(\beta)/\beta$ on the RHS of the equation.  
Riemann \cite{riemann} denotes this function by $f(x)$ and 
Edwards by $J(x)$ \cite{edwards}. The number of primes less than 
$x$, denoted by $\pi(x)$, may be expressed in terms of $f(x)$ as 
\begin{equation} \label{pix}
\pi(x)=\sum_{n=1}^{\infty} {\mu(n) f(x^{1/n})\over n}~,
\label{riesel2}
\end{equation} 
where $\mu(n)$ is the M\"{o}bius function~\cite{edwards}. The modulating 
effect of the oscillatory terms due to the first twenty nine pairs of the 
complex Riemann zeros was numerically examined by Riesel and G\"{o}hl 
~\cite{riesel} in 1970. This early work already showed the approximate 
formation of 
the first few steps at the prime numbers, and modulations for some larger 
primes. Note that the series (\ref{riesel2}) requires a knowledge of the 
M\"{o}bius function, and is much more complicated than Eq.~(\ref{best2}). 
Riesel and G\"{o}hl actually replaced the sum over the M\"{o}bius functions 
by the Gram series involving factorials which are difficult  
to compute accurately for large integers.  In this paper, we shall rather 
study formula (\ref{best2}) for $\rho(x)$ since it contains more information 
than the formula for the density of the primes (no powers)  
which can be obtained from the derivative of $\pi(x)$.

\section{Numerics}

Numerically, we can only evaluate a finite number of terms from 
Riemann's infinite series. Although it would seem by inspection that all 
terms of the series are equally important and that there is no optimal  
ordering of the terms, Riemann states that the series is conditionally 
convergent and that it must be summed in order of increasing size of 
$\alpha$. 
(For any series whose convergence is conditional, the order of summation 
must be specified since different orderings produce different results.)
In this case, the ``natural order'' is the correct one. Riemann goes on 
to state that with this ordering the truncated series should give an 
approximation to the density of primes (and their powers), but that using 
a different ordering, the resulting finite series can approach arbitrary 
real values. We have verified this numerically and found that using  
finite sets of zeros chosen according to different rules yields  
incorrect results. Thus, for the numerical work that follows, we use the  
correct ordering. 

\subsection{Line Intensities of the Truncated Series}
   
\begin{figure} 
\scalebox{0.453}{\includegraphics*{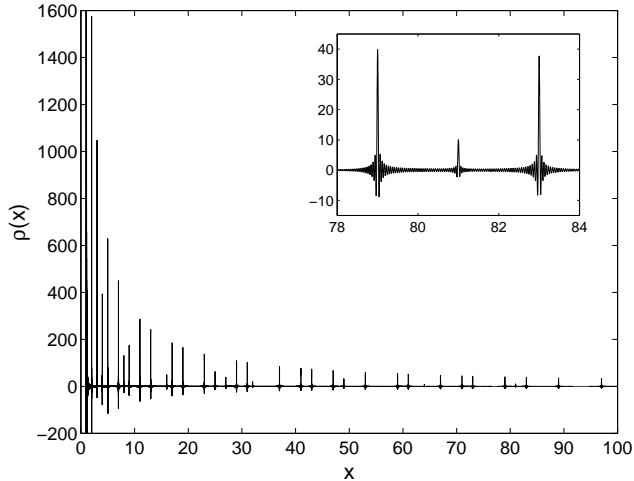}}
\caption{\label{raw} The result of computing Eq.~(\ref{best2}) 
using the first $10^4$ zeros for $x \in [1.5,100]$. The inset 
shows a closer view of three lines that appear for 
$x \in [77,84]$. The three lineshapes are similar so that their relative 
heights are somewhat meaningful. However, the lineshapes 
vary considerably throughout the entire range so that  
heights cannot be immediately interpreted as relative intensities.}
\end{figure}

\begin{figure} 
\scalebox{0.453}{\includegraphics*{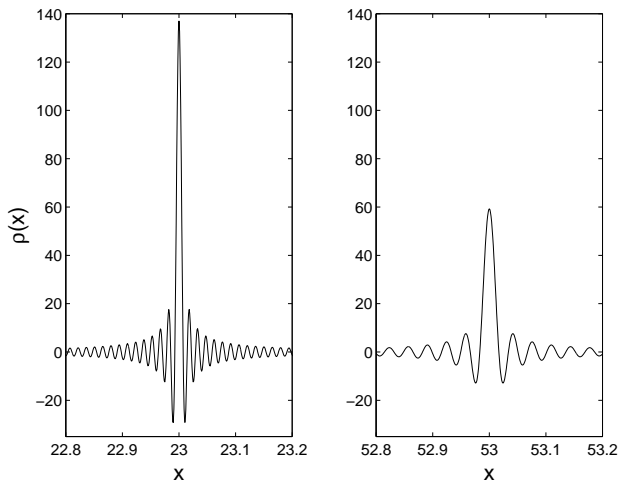}}
\caption{\label{closer} A closer view of two non-adjacent lines in 
Fig.~\ref{raw}. 
The lineshapes clearly differ so that their relative heights are not 
meaningful.}
\end{figure}

We first computed Eq.~(\ref{best2})  
using the first $10^4$ zeros and observed lines at the positions of the 
primes and their powers for $x<5000$. 
However, for $x>2000$, many lines cannot be fully resolved and the signal 
eventually dies out.  
This is due to truncation since only a small number of 
zeros have been included. (This will be discussed in more detail below.) 
Nevertheless, even this small number of zeros yields narrow lines at 
the lowest primes. 
In Fig.~\ref{raw}, we display the result for $x \in [1.5,100]$. 
Although one can clearly observe lines at the positions of the primes, 
the relative intensities cannot be determined by inspection since   
the lineshapes are not uniform (see Fig.~\ref{closer}). 
This is a common problem in spectral analysis, and is often resolved 
by imposing a more uniform 
lineshape through convolution of the signal with an appropriate smooth 
``response function'' \cite{NR}.  
The response function is typically a peaked 
function that falls to zero in both directions from its maximum.  
Gaussian functions   
are positive-definite and decay rapidly.  
They are also convenient to use since their shape only depends on a single 
parameter (the variance) and therefore can be easily controlled.    

Thus, we next convolve this approximate density $\rho(x)$ 
(i.e. smooth term and truncated series) with an 
\emph{unnormalized} Gaussian of variance $\sigma$:
\be \label{rhosmt}
\rho(x) *
G_{\sigma}(x) = \int_{1}^{\infty} \rho(x') G_{\sigma}(x - x') \d x',
\ee
where 
\be \label{Gsmth}
G_{\sigma}(x) = \exp(-x^2/{2 \sigma^2}).
\ee
The effect of the convolution is that  
rapidly oscillating features are washed out and smooth peaked features   
are smeared into the shape of the response function. If the lines were 
perfect $\delta$ functions of height $D_n$, then from Eq.~(\ref{rhosmt}), 
these
would be replaced by $D_n G_{\sigma}(x-x_n)$, i.e. Gaussians of variance 
$\sigma$ with height $D_n$ at $x=x_n$.
Of course, the lines are not $\delta$-function spikes 
so that the resultant lineshapes are not exactly Gaussian, but as long as the 
intrinsic linewidth is sufficiently small compared to the variance, 
the deviation from a perfect Gaussian is quite negligible. 
Therefore, the convolution  
produces a series of Gaussian lines, each of the same width.  
The key point is that the lineshapes are now essentially uniform so 
that the actual heights can be meaningfully compared and 
immediately interpreted as the relative intensities.
It is important to 
keep in mind that since the response function has a maximum height of 
unity, the height of a line \textit{after} convolution should be 
the area under that line \textit{before} convolution. 
The reason for this is that 
although the lines of the original signal act like $\delta$ functions with 
respect to the response function, they do have nonzero widths and so their 
effective $\delta$-function ``heights'' $D_n$, are equal to the areas. 
In this sense, the 
convolution procedure is equivalent to directly integrating the area under 
each line of the signal. However, the convolution technique is much 
simpler and avoids errors that can arise from the long oscillatory tails of 
individual lines.   
Note that this procedure cannot resolve two adjacent lines when the spacing 
between them is smaller than $\sigma$ and thus $\sigma_{\text{max}}=1/2$.

\begin{figure} 
\scalebox{0.453}{\includegraphics*{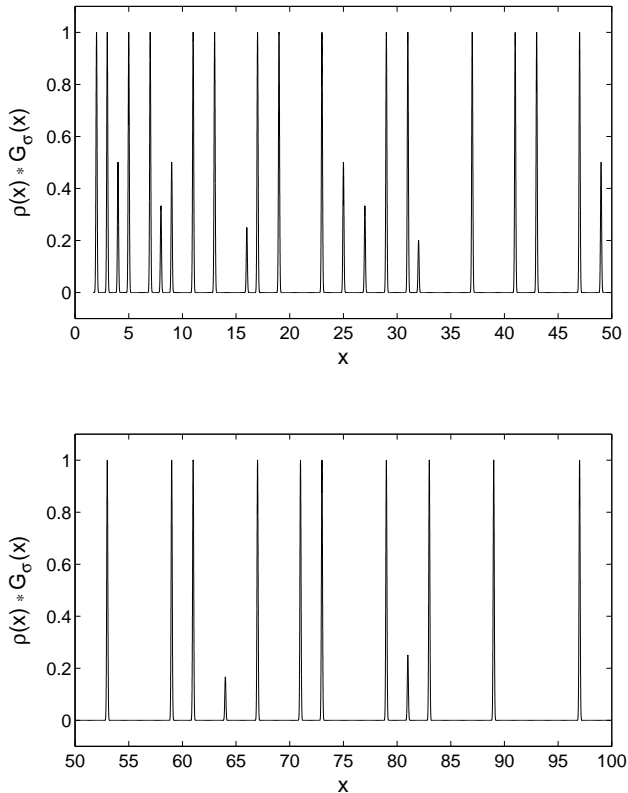}}
\caption{\label{true} The result of computing Eq.~(\ref{rhosmt}) 
using the first $10^4$ zeros and $\sigma=0.05$. The range is the same as in 
Fig.~\ref{raw}. Note that lines with height less than unity occur at 
powers of primes $p^n$ and have height $1/n$ (for example, the  
line at $2^6=64$ has height $1/6=0.1\dot{6}$).}
\end{figure}

We have computed Eq.~(\ref{rhosmt}) for the range of interest in 
Fig.~\ref{raw}. This is shown in Fig.~\ref{true} using $\sigma=0.05$. 
We note here that the heights do not depend on the specific value 
of $\sigma$ due to the fact the Guassian is not normalized. 
As more terms are included in the sum, the natural linewidths decrease and the 
convolution becomes more accurate. It is then possible to produce 
high-resolution 
lines by using smaller variances. For example, using $10^5$ zeros, 
we produced lines with a variance $\sigma=0.01$. 
It would be useful to know how many primes can be resolved using a 
prescribed number of zeros. In the present scheme, one simply observes 
where the lines of the original signal develop a sufficiently large width. 
The important criterion here is that all linewidths should be at least 
smaller than the mean spacing between all powers of primes in the interval 
of interest. 
Of course, the width of any line is related to the number of terms used 
in truncating the series.
Although this relationship can be determined, there is  
still the problem that all the lines have different shapes.     
Thus, we shall find it more useful to determine for what values of $x$ 
the truncated formula can no longer produce lines of a specific 
uniform shape.  

\subsection{Sum Rule and Numerical Convergence}

The convergence of the series can be examined using a more 
controlled application of the 
convolution procedure described above.
The general idea is to construct a series (``sum rule'') 
that absolutely converges to a ``coarse-grained'' version of the exact 
density. This density is obtained by replacing all spikes of the exact 
density by smooth peaked functions. One immediate advantage is  
improved convergence since it is easier to reproduce the well-defined 
\emph{smooth} peaked functions of a coarse-grained density using a 
truncated sum rule than it is to reproduce spikes using the original 
truncated series. However, for our purposes, the more important reason  
for using a sum rule is to \emph{control} the convergence of the series. 
This will become evident after the sum rule is given.
The sum rule itself is obtained from a direct convolution of the original 
series with a ``smoothing function'', that is, some smooth function whose 
Fourier coefficients rapidly decrease. (Since the original series consists 
of cosines, the resulting integral is essentially a 
cosine transform of the smoothing function.) 

The above discussion is quite general. We now connect this idea with the  
numerical calculations described above. Assume 
the coarse-grained density consists of a set of Gaussian 
functions of variance $\sigma$ centered at each prime (or prime power) with 
heights equal to unity (or the reciprocal power). We want to construct a 
series as described above that converges to this density, that is, 
we want to find a Gaussian sum rule for Riemann's series. 
To do this, we convolve Riemann's series term-by-term 
with a Gaussian smoothing function. 
For the following calculation, we will define $S_\alpha(x)=\alpha \ln x$ 
and $A(x)=-2/\sqrt{x}\ln x$. Then, we can 
write Riemann's formula as $\tilde{\rho} (x) = A(x) \sum_\alpha \text{Re} 
\{ \exp [i S_\alpha(x)] \}$. The Gaussian sum rule is    
\bea \label{sumrule1}
\tilde{\rho}_\sigma (x)&=& \tilde{\rho} (x) * G_\sigma(x)=
 \int_{-\infty}^{\infty} 
\tilde{\rho} (x') e^{-(x-x')^2/2\sigma^2} \d x' \nonumber \\
&=& \sum_\alpha \text{Re} \left \{ \int_{-\infty}^{\infty} 
A(x') e^{i S_\alpha(x')} e^{-(x-x')^2/2\sigma^2} \d x' \right \}, 
\quad \quad x>1.
\eea
For $\sigma \le \sigma_{\text{max}}=1/2$, the Gaussian rapidly decays 
to zero. This implies that 
the main contribution to the integral comes from a small interval 
centered about $x'=x$. Elsewhere, the integrand is practically zero. Thus, 
we make two approximations to proceed further. First, the 
amplitude function $A(x')$ changes 
very slowly and on the small interval of interest 
$A(x') \approx A(x)$. Secondly, the phase function $S_\alpha(x')$ can be 
replaced by its Taylor series expansion about $x'=x$: 
$S_\alpha(x')=S_\alpha(x)+S'_\alpha(x)(x'-x) + \ldots$. 
If we retain the leading-order term only,    
\bea \label{sumrule2}
\tilde{\rho}_\sigma (x)
&\approx& A(x) \sum_\alpha \text{Re} \left \{ \int_{-\infty}^{\infty} 
e^{i [S_\alpha(x)+S'_\alpha(x)(x'-x)]} e^{-(x-x')^2/2\sigma^2} \d x' 
\right \} \nonumber \\
&=& A(x) \sum_\alpha \text{Re}  \left \{ e^{i [S_\alpha(x)-xS'_\alpha(x)]}
e^{-x^2/2\sigma^2} \int_{-\infty}^{\infty}
e^{-[x'^2-(2x+2i\sigma^2S'_\alpha(x))x']/2\sigma^2} \d x' \right \}  
\nonumber \\
&=& \sqrt{2\pi}\sigma A(x)\sum_\alpha  
e^{-\sigma^2S'^2_\alpha(x)/2} \text{Re}  \left \{e^{iS_\alpha(x)} \right \},
\eea
where we have used the standard result for the Gaussian integral 
\cite{integral}. Finally, the Gaussian sum rule for Riemann's series is 
\bea \label{rulefin}
\tilde{\rho}_\sigma (x)=- {{2\sqrt{2\pi}\sigma} \over {\sqrt{x}\ln x}} 
\sum_\alpha e^{-\sigma^2\alpha^2/2x^2} \cos (\alpha \ln x).
\eea
This sum rule explicitly shows the effect of convolution on the series; 
each term is modulated by an exponential factor. This factor essentially 
controls the convergence of the series for all values of $x$. Although 
the orginal series is only conditionally convergent, as long as the correct 
ordering is used, this sum rule is also \emph{absolutely} convergent. 
As stated above, we seek an approximate relation between the maximum  
zero included in the sum and the maximum prime that can be resolved. 
One way to determine this is as follows. First, specify the 
value of the largest zero, $\alpha_{\text{max}}$ and include all zeros 
$\alpha \le \alpha_{\text{max}}$. Then, there exists a set of values 
$x < x_{\text{max}}$ for 
which the exponential factor falls below some threshold parameter 
$\varepsilon$. This condition immediately gives the simple relation 

\be \label{converge1}
x_{\text{max}}=\left[{{\sigma} \over {\sqrt{-2 \ln (\varepsilon)}}} \right]
\alpha_{\text{max}},
\ee
where $0<\varepsilon<M$.  For $\alpha > \alpha_{\text{max}}$ and 
$x \le x_{\text{max}}$, all terms are  
exponentially smaller than $\varepsilon$ and are thus 
numerically insignificant. 
The choice of the parameter $\varepsilon$ depends on the desired 
precision of a resolved line. An upper bound 
$M$ for the parameter is the value of the exponential factor 
($e^{-3/2}$) at its inflection point $x_I=(1/\sqrt{3})\sigma \alpha$ 
\cite{inflectsmall}. This implies $x_{\text{max}}<x_I$. 
The lower bound can be as small as machine zero  
(for example $10^{-16}$). However, there is no reason for such an extreme 
choice since we are mostly interested in determining where numerical errors 
become significant (i.e. where lines are no longer \emph{visibly} resolved 
and the intensities are erroneous by more than 1\%).  
Of course, higher precision can be imposed at the cost of resolving fewer 
primes. But, since the improved precision will not be apparent in the 
graph of $\rho_\sigma (x)$, there is no compelling reason to choose 
exceedingly small values. For our purposes, a convenient choice is 
$\varepsilon=e^{-7/2}$. 

\begin{figure} 
\scalebox{0.45}{\includegraphics*{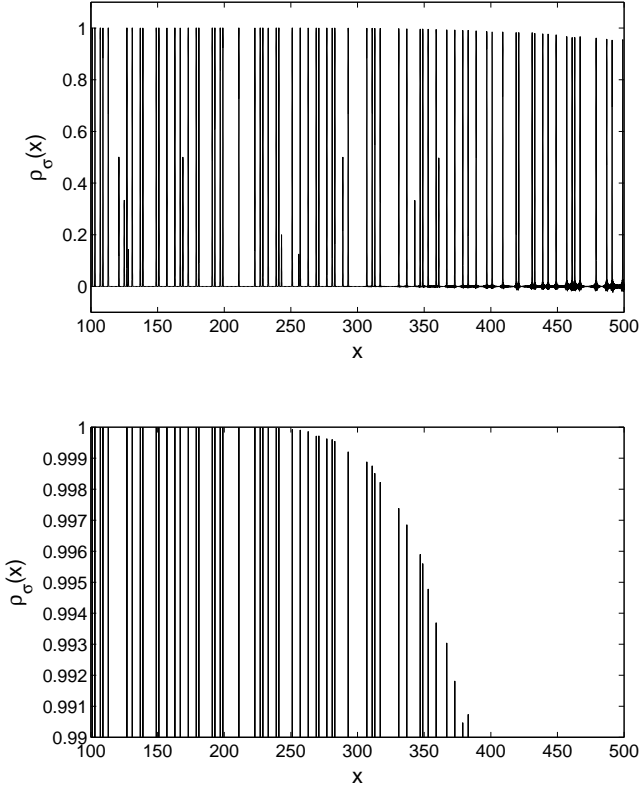}}
\caption{\label{tentzeros} The sum rule (\ref{rulefin}) 
using the first $10^4$ zeros and $\sigma=0.1$. The lower window shows
a closer view of the lines in the interval $0.99<\rho_\sigma(x)<1$. 
Formula (\ref{converge1}) 
indicates that for $x \le x_{\text{max}} \doteq 373$, all lines should be 
resolved and intensities should have errors less than $\sigma^2=0.01$.}
\end{figure}

\begin{figure} 
\scalebox{0.45}{\includegraphics*{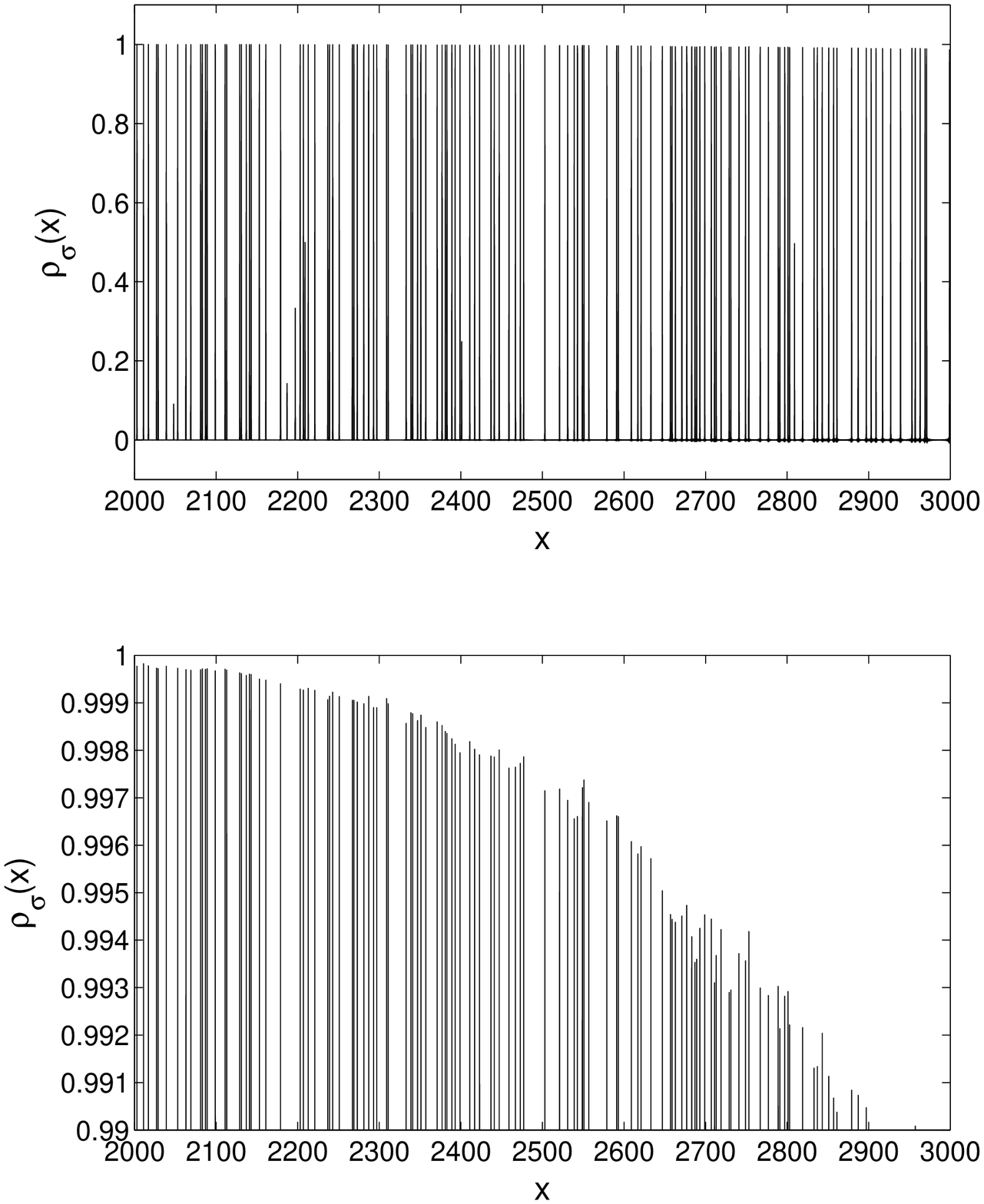}}
\caption{\label{hundredtzeros} The sum rule (\ref{rulefin}) 
using the first $10^5$ zeros and $\sigma=0.1$. The lower window shows
a closer view of the lines in the interval $0.99<\rho_\sigma(x)<1$. 
Formula (\ref{converge1}) 
indicates that for $x \le x_{\text{max}} \doteq 2832$, all lines should be 
resolved and intensities should have errors less than $\sigma^2=0.01$.}
\end{figure}

We now provide a few examples to illustrate the utility of relation 
(\ref{converge1}). As a first example, we take $\alpha_{\text{max}} \doteq
9878$ which is the $10^4$th zero. Using the above 
formula (with $\sigma=0.1$) yields $x_{\text{max}} \doteq 373$. 
In Fig.~\ref{tentzeros}, we evaluate Eq.~(\ref{rulefin}) using the first 
$10^4$ zeros (and include the smooth term). One 
can clearly see significant errors for $x > 400$. As a second example, we 
take $\alpha_{\text{max}} \doteq 74921$ (the 
$10^5$th zero). The formula then gives $x_{\text{max}} \doteq 2832$. 
In Fig.~\ref{hundredtzeros}, we truncate
Eq.~(\ref{rulefin}) at this value of $\alpha_{\text{max}}$ and 
observe significant errors occur for $x > 2900$. 

An additional benefit of the sum rule is that it gives us an 
immediate measure of 
the error incurred from truncation. The largest errors are in the vicinity 
of $x_{\text{max}}$ where there are contributions $O(\varepsilon)$ that 
have been excluded. For all other values of $x<x_{\text{max}}$, 
the excluded terms are exponentially smaller. Of course, we have complete 
control of this error through our freedom in specifying $\varepsilon$. 
In the case of the original truncated series, it is not 
immediately obvious what the errors are, but they can be determined through 
more elaborate analysis.   

\section{Discussion and Conclusion}

By writing $\zeta(\gamma+it)=|\zeta(\gamma+i t)|
\exp (-i\theta_{\gamma}(t))$, 
we see that all the information about the zeros along the $t-$axis is 
contained in the phase $\theta_{\gamma}(t)$. This has to jump by $\pi$ to 
accomodate the sign change in $\zeta$ at every zero, and it can be 
shown that the 
oscillating part of the density of the zeros on the critical line is 
proportional to the derivative of the imaginry part of 
$\ln \zeta(t)$ with respect to $t$ \cite{brack}. 
On the other hand, we see from Eq.~(\ref{best1}) that 
the appropriate contour integral over $\ln \zeta(\beta)$ also yields 
$\rho(x)$ relating to the primes. Thus, the phase of the zeta function, as 
defined above, connects the Riemann zeros to the primes.  

As mentioned above, if the series is truncated, the signal gradually 
dies out as $x$ increases. This can be understood by noting that due to the  
logarithmic dependence, each term produces an oscillation whose 
period continually increases while its amplitude decays. Clearly, 
more high frequency (large 
$\alpha$) terms are required for sufficient constructive interference. 
This explains the fact that lines at small values of $x$ are resolved more 
quickly than at larger values. Although  
the higher frequency terms are responsible for short-range oscillations and 
one could imagine exclusive use of those terms rather than lower frequency 
terms, the difficulty 
is the conditional convergence of the series and the fact that all of the 
terms are equally important. Unfortunately, this implies that Riemann's 
formula is impractical for resolving lines at large primes. This is also 
consistent with Eq.~(\ref{converge1}). If one is interested in 
using Riemann's series to find (new) large primes, for example, on the 
order of $10^{250,000}$, then one requires an accurate knowledge of 
roughly the same number of zeros. 

We emphasize that Riemann's formula, is 
only correct if the Riemann hypothesis is true. Otherwise,  
if a pair of zeros occur at $\beta_{\pm}=\gamma \pm i\alpha$, the 
factor $x^{1/2}$ in the denominator of the oscillating term of 
Eq.~(\ref{best2}) should be replaced by $x^{(1-\gamma)}$ \cite{keating}.
An interesting numerical experiment is to move the zeros off the critical 
line, that is, to arbitrarily change their real parts. 
We find this still produces lines at the primes and their powers, 
but the relative intensities are incorrect. This is interesting since 
it demonstrates that the location of the primes depends only on the 
imaginary part of the zero. The real part only affects the 
intensities which are anyway not evident from a direct evaluation 
of the series. This provides another motivation for the numerical and 
analytical procedures described in this paper.   

It is natural to compare the oscillating part of 
the density $\tilde{\rho}(x)$ with the semiclassical trace 
formula \cite{gutz2,gutzwiller} of a dynamical system. 
One could identify $\alpha$ as an orbit label, 
one for each zero of the zeta function and  
$x$ as the single-particle energy variable. Then, $\rho(x)$ in 
Eq.~(\ref{best2}) may be interpreted as the density of states as a 
function of energy with the 
first term on the RHS corresponding to the smooth 
Thomas-Fermi (TF) contribution~\cite{note1}. 
In the oscillating part, the argument $\alpha \ln x$ of the cosine term 
should then correspond to the action $S_\alpha(x)$ of the orbit $\alpha$. 
Note however that there are no implicit repetition indices in 
Eq.~(\ref{best2}) thereby implying that even if one gives a dynamical  
interpretation to $\tilde{\rho}(x)$, the orbits are not periodic.   
This is in direct contrast to the trace formula for the Riemann zeros, in 
which the orbits are periodic with primitive period $\ln p$ for each prime 
\cite{berry}. 
Of course, the most striking feature is that 
the amplitude has no $\alpha$ dependence. Even oscillatory contributions 
to the density of states from non-periodic trajectories usually have 
amplitudes that depend on the orbit \cite{nonPOs}.     
In the event that there is a fortuitous cancellation of the index $\alpha$,
it is unlikely that the energy dependence in the denominator of the 
oscillating term as well as the TF term can then be generated 
consistently by the same Hamiltonian. Consequently, Riemann's formula is not 
a trace formula of dynamical origin. 

With regard to spectral statistics,   
it is well known that nearest-neighbour spacings (NNS) \cite{bohigas} of 
the Riemann zeros obey the GUE distribution of Random Matrix Theory, 
characteristic of a chaotic quantum system without time-reversal 
symmetry \cite{odlyzko,mehta}. The same zeros also generate 
the spectrum of the primes and their powers through Riemann's formula 
(\ref{best2}). As mentioned earlier, the  NNS distribution 
of the primes is Poisson-like~\cite{porter}, with  
some level repulsion, which, if at all of dynamic origin, hints only to 
near-integrability \cite{stockmann}. Thus, it is quite remarkable that 
the highly-correlated sequence of the zeros can (through Riemann's formula) 
interfere to produce the almost-uncorrelated sequence of the primes.

In conclusion, we have demonstrated that the spectrum of the primes and their 
integer powers can be accurately generated from a sum of periodic terms, 
each term involving a zero of the zeta function. This is in the spirit 
of semiclassical periodic orbit theory, where the individual levels of 
a quantum spectrum may be resolved from a sum of oscillatory terms, 
each arising from periodic orbits. 
Despite the accuracy of the generated spectrum, Riemann's formula is not a 
trace formula. However, this does not imply that there is no such formula, 
and it would still be interesting  
to understand the spectrum of the primes in terms of  
periodic orbits. This could provide insight into the structure of a 
possible trace formula for the primes. 
If this formula could be found, 
the remaining challenge would be to obtain the corresponding Hamiltonian.

\begin{acknowledgments}
We acknowledge Ranjan Bhaduri, Randy Dumont, Avinash Khare, 
John Nieminen, Muoi Tran, Jim Waddington for useful discussions, 
and Andrew Odlyzko for supplying the Riemann zeros.
This work was financially supported by the Natural Sciences and Engineering Research Council of Canada (NSERC).
\end{acknowledgments}


\begin{thebibliography}{99}
\bibitem{integral} $\int_{-\infty}^{\infty}e^{-ax'^2+bx'}\d x' = 
\sqrt{\pi \over a} e^{b^2/4a} \quad (a>0)$. 
It is not necessary to evaluate the 
convolution integral over the interval $(1,\infty)$. 
If one insisted on doing so, the integral at the end is rather 
$\int_{1}^{\infty}e^{-ax'^2+bx'}\d x' = \sqrt{\pi \over 4a} e^{b^2/4a} 
\text{erfc}\left(-{b \over 2 \sqrt{a}}+\sqrt{a}\right)$. The resulting 
sum rule is slightly more complicated, but the difference is insignificant 
since $\sigma$ is small.
\bibitem{inflectsmall} For smaller values of $x$, a more accurate 
upper bound is given by the inflection point of 
$x^{-1/2}e^{-\sigma^2\alpha^2/2x^2}$ which is 
${\sqrt{2(12-3\sqrt{13})} \over 3}\sigma\alpha$. If the logarithmic 
term is also 
included, the inflection points are roots of a transcendental equation.
\bibitem{note1} The second smooth term on the RHS of Eq.~(\ref{best2}) 
arises from the trivial zeros, and may not be taken as part of a TF term.
\end{thebibliography}
\end{document}